\documentclass[12pt,preprint]{aastex}

\shorttitle{VLA and XMM-Newton observations of W41/HESS J1834-087}
\shortauthors{Tian, Li, Leahy \& Wang}

\begin{document}

\title{VLA and XMM-Newton observations of the SNR W41/TeV $\gamma$-ray source HESS J1834-087}

\author{W.W. Tian\altaffilmark{1,2}, Z. Li\altaffilmark{3}, D.A. Leahy\altaffilmark{2}, Q.D. Wang\altaffilmark{3}}
\altaffiltext{1}{National Astronomical Observatories, CAS, Beijing 100012, China; email: tww@iras.ucalgary.ca} 
\altaffiltext{2}{Department of Physics \& Astronomy, University of Calgary, Calgary, Alberta T2N 1N4, Canada}
\altaffiltext{3}{Department of Astronomy, University of Massachusetts, 710 North Pleasant Street, Amherst, MA 01003, USA}

\begin{abstract}
The recently discovered extended TeV source HESS J1834-087 is 
associated with both a diffuse X-ray enhancement and a molecular cloud,
projected at the center of an old radio supernova remnant G23.3-0.3 (SNR W41). New HI observations from the VLA Galactic Plane Survey (VGPS) show unambiguous structures associated with W41 in the radial velocity range of 53 to 63 km/s, so we obtain for W41 a distance of 4$\pm$0.2 kpc. A new higher sensitivity VGPS continuum image of W41 at 1420 MHz shows faint emission in its eastern part not detected by previous observations, so we provide a new angular size of 36$^{\prime}$$\times$30$^{\prime}$ in $b$-$l$ direction (average radius of 19 pc). We estimate for W41 an age of $\sim$10$^{5}$ yr. 
A new XMM-Newton observation reveals diffuse X-ray emission within the HESS source and suggests an association between the X-ray and $\gamma$-ray emission. The high-resolution $^{13}$CO images of W41 further reveal a giant molecular cloud (GMC) located at the center of W41, likely associated with W41 in the the radial velocity range of 61 to 66 km/s. Altogether, the new observations can be interpreted as providing the first evidence that an old SNR encounters a GMC to emit TeV gamma-rays in the cloud material. 
\end{abstract}


\section{Introduction}
Young supernova remnants (SNRs) are one of the main galactic populations to generate very high energy (VHE, above 10$^{11}$ eV) $\gamma$-rays (Torres et al. 2003). Observationally, TeV $\gamma$-rays have been detected from the Crab Nebula (SN1054, Weekes et al. 1989), the SNRs RX J1713.7-3946 (G347.3-0.5, Enomoto et al. 2002) and RX J0852.0-4622 (G266.2-1.2, Aharonian et al. 2005a). Theoretically, acceleration mechanisms for relativistic electrons at SNR shock fronts have been well-established (Malkov $\&$ Drury 2001).
Recently, Aharonian et al.'s (2006) survey of the inner part of our Galaxy has revealed 14 new TeV $\gamma$-ray sources. The origin of some of them remainis uncertain. Yamazaki et al. (2006) showed that TeV $\gamma$-rays can originate from an old SNR of an age around 10$^{5}$ yr or from a giant molecular cloud (GMC) encountered by the SNR,
via pion decay from proton-proton collisions. In their scenario, the flux ratio of the $\gamma$-rays to the associated X-ray emission is much higher than that from a young SNR. 
In this letter, we provide observational evidence supporting that an old SNR encounters a GMC to emit TeV $\gamma$-rays, based on new radio and X-ray observations as well as recent $^{13}$CO images of SNR W41, which is spatially coincident with HESS J1834-087, one of the 14 TeV sources.

\section{Radio and X-ray Observations}
The radio continuum and HI emission data sets come from the Very Large Array (VLA) Galactic Plane Survey (VGPS), described in detail by Stil et al. (2006). The data sets are mainly based on observations from VLA of the National Radio Astronomy Observatory (NRAO). The spatial resolution of the continuum images of W41 is 1$^{\prime}$ (FWHM) at 1420 MHz. The synthesized beam for the HI line images is 1$^{\prime}$ and the radial velocity resolution is 1.56 km$/$s.  The short-spacing information for the H I spectral line images is from additional observations with the 100 m Green Bank Telescope of the NRAO. 

W41 was observed by {\sl XMM-Newton} on September 18, 2005 (Obs.~ID 0302560301; PI: G. Puehlhofer), with
a 20 ksec exposure. In this work, we only used data obtained from the EPIC-PN. We used SAS, version 7.0.0, for data reduction. We
selected PN events with patterns 0 through 4 and applied flag filter FLAG==0. Excluding time intervals contaminated
by background flares results in a net exposure of 12.4 ksec. We then constructed exposure maps in the 0.3-0.7, 0.7-1.5,
1.5-3, and 3-7 keV bands for flat-fielding. We applied the ``filter wheel closed" (FWC) data for 
instrumental background subtraction. We also searched for point-like sources using a detection procedure detailed by Wang (2004).

\section{Results}
\subsection{Continuum Emission}

The VGPS continuum image of W41 at 1420 MHz is shown in Fig. 1. The VGPS map has a higher resolution (by a factor of 3) and sensitivity, and shows more details than the previous image at 330 MHz (Kassim 1992). A prominent filament structure outlines W41. Fainter features, not detected by previous observations, appear in its eastern part. The new image gives a corrected angular size of W41: 36$^{\prime}$$\times$30$^{\prime}$ in $b$-$l$ direction. The HII regions overlapping in W41 have been resolved into at least three components.
We have derived an integrated flux density of 59.7$\pm$8.2 Jy for W41 (including HII regions in the SNR) at 1420 MHz. The resulting 330-1420 MHz spectral index (the flux density is 143 $\pm$29 Jy at 330 MHz including HII regions, Kassim 1992), has a lower limit of 0.43 ($S_{\nu}\propto\nu^{-\alpha}$). The TeV $\gamma$-ray source HESS J1934-087, detected by Aharonian et al.(2005b), is located at the center of W41. Its location and extent are shown by a
solid circle (white), centered at ({\sl l}, {\sl b})=(23.24, -0.32) with a radius of 5\farcm4 (Aharonian et al.~2006). The pulsar PSR J1933-0827 also shown in Fig. 1 was proposed to be associated with W41 by Gaensler and Johnston (1995).

\subsection{HI and CO Emission}
We have searched the VGPS radial velocity range for features in the HI which might be related to the morphology of W41. There is unambiguous HI emission coincident with the SNR in the velocity range: 53 to 63 km/s. Fig. 2  is the column density map of HI emission integrating over channels from 53 to 63 km/s in units of 10$^{20}$ atom cm$^{-2}$. The map has superimposed the 30 K contour of 1420 MHz continuum emission chosen to show the SNR. The circle is the same as in Fig.1.  

We extract $^{13}$CO images of W41 from the survey of Jackson et al.(2006). A giant molecular cloud is found at the center of W41 and in the radial velocity range of 61 to 66 km/s so is highly likely associated with W41. Fig. 3 shows the averaged $^{13}$CO map for channels from 61 to 66 km/s.  The contour and circle in the map are the same as in Fig. 2.

\subsection{X-ray Emission}

The X-ray images of W41 are shown in Fig.~\ref{fig:w41_x} for the soft (0.3-1.5 keV) and hard
(3-7 keV) bands. In the soft band, the bulk of the field is of smooth, low intensity. 
In the hard band, a region of enhanced intensities is clearly present within
the extent of HESS J1834-087. With the resolution of the instrument (FWHM $\sim 13^{\prime\prime}$), the enhancement is apparently extented by visual comparison with nearby point sources in the range, although we cannot completely rule out the possibility that it represents an unusual cluster of point-like sources. Given its location, this feature is likely associated with HESS J1834-087. 

We perform spectral analysis for the feature, for which
we extract a spectrum from a circle centering at ({\sl l}, {\sl b})=(23.243, -0.331)
with a radius of 2\farcm5 (Fig.~\ref{fig:w41_x}).
To determine the local sky background, we extract a spectrum from a concentric circle with a radius of 7\farcm5
and with the enclosed circle representing HESS J1834-087 and detected point-like sources excluded.
The two spectra are shown in Fig.~\ref{fig:w41_spec}. Part of the background emission may arise from
the interior of the remnant and vary between the source and background regions . Thus our procedure of background
determination might be affected by this non-uniformity. Nevertheless, 
encouraged by the apparently uniform surface intensity at energies below 1.5 keV (Fig.~\ref{fig:w41_x}),
we assume that the background emission also varies little at higher energies.
We use the X-ray Spectral Fitting Package (XSPEC) to fit the background spectrum and find that it can be characterized by a combined model consisting of a thermal plasma component 
(APEC in XSPEC) and a power-law component, both subject to absorption (Table~\ref{tab:spec}).
These two components, scaled accordingly to the sky area, are applied to account for the background contribution
in the source spectrum. The remaining emission in the source spectrum, presumably intrinsic to the feature,
is then characterized by an additional component, for which we find a heavily absorbed power-law.
Fitting results are listed in Table~\ref{tab:spec}.

\section{Discussion and Conclusion}  
\subsection{Distance and Age of W41}
Using flat galactic rotation velocity V$_{R}$=V$_{0}$=200 km/s and R$_{0}$=8.0 kpc,  we obtain the SNR's distance of 4$\pm$0.2 kpc or 10.7$\pm$0.2 kpc.  The updated average angular diameter of W41, 33$^{\prime}$, yields radius about R=19 pc (d=4 kpc) or 51 pc (d=10.7 kpc) for the SNR. Since known shell-type supernova remnants are a few to $\sim$20 pc in radius, we adopt the closer distance. 
From the HI column density map associated with W41 (Fig. 2), we estimate a N$_{H}$ of about 5 - 10 $\times$$10^{20}cm^{-2}$, so the density n$_0$=N$_{H}$/(2R) around W41 is about 6 $cm^{-3}$.  Applying a Sedov model (Cox, 1972), for a typical explosion energy of E=0.75$\times$ $10^{51}$ erg, yields an age of $\sim$6$\times$ $10^{4}$ yr. However, for n$_{o}$$\sim$6 $cm^{-3}$ one finds R $>$ R$^{(c)}_{s}$ (the radius for complete cooling; Cox, 1972). The age determined from the complete cooling expansion is $\sim$2$\times$ $10^{5}$ yr.  
 
\subsection{X-ray, TeV $\gamma$-ray and CO Emission from W41}
Previous $^{12}$CO observations have shown that W41 is associated with a very large molecular complex (Dame et al. 1986). Albert et al. (2006) studied the $^{12}$CO images from Dame et al (2001) and suggested the giant molecular cloud associated with W41 is best defined by integrated the $^{12}$CO peak emission from 70 to 85 km/s. They also used the $^{13}$CO images from Jackson et al. (2006) and confirmed that the  recently discovered TeV source HESS J1834-087 lies towards a GMC. However, our observations show the HI-line emissions are associated with W41 in the velocity range of 53 to 63 km/s. $^{13}$CO is useful as optically thin tracer of the molecular cloud so we reanalyzed the $^{13}$CO images of W41. We found a GMC located at the center of W41 in the radial velocity range of 61 to 66 km/s. Fig. 3 shows the bright $^{13}$CO emission in the velocity range and that it is coincident with HESS J1834-087. The total $H_{2}$ mass of the CO emission peak (over 0.1$^{o}$$\times$0.2$^{o}$ region) is estimated from $M_{H_{2}}$ = $N_{H_{2}}$$\Omega$$d^{2}$(2m$_{H}$/$M_{\odot}$). We take $N_{H_{2}}$/$W_{CO}$$\approx$1.8$\times$10$^{20}$ cm$^{-2}$ K$^{-1}$ km$^{-1}$s from Dame et al. (2001). The total integrated intensity of $^{13}$CO is $W_{CO}$ $\approx$ 5 K km/s from Fig. 3. Assuming a $^{12}$CO/$^{13}$CO isotopic abundance ratio of 30 (Langer \& Penzias 1990), we obtain an average $H_{2}$ column density of $N_{H_{2}}$$\approx$2.7$\times$10$^{22}$ cm$^{-2}$, and a molecular cloud mass of $M_{H_{2}}$$\approx$4.5$\times$$10^{4}$$M_{\odot}$. This is a giant molecular cloud with a density of $\sim$10$^{3}$cm$^{-3}$. 

From the observed $\gamma$-ray luminosity (Albert et al. 2006), using equation 16 in Torres et al. (2003) and a supernova power of 10$^{51}$ ergs, we obtain a relation between an acceleration efficiency $\theta$ of hadrons and the required density $n$ of matter in the $\gamma$-ray production region for hadronic origin of the observed radiation:       
$\theta$$\sim$10$^{-2}$ requires $n$$\sim$10$^{2}$cm$^{-3}$; $\theta$$\sim$10$^{-3}$ requires $n\sim$10$^{3}$cm$^{-3}$ (similar values are obtained by Combi et al. for another SNR likely interacting with a massive cloud, 1998). Generally, the maximum acceleration efficiency $\theta$ of 10\% is accepted, so the GMC is dense enough to produce the observed TeV intensity with a lower acceleration efficiency. 
  The protons could diffuse only partway into the cloud, resulting in the observed offset of peak CO and p-p to $\pi$$^{0}$ decay TeV $\gamma$-rays. The physical peak of X-ray and TeV should be coincident. The column density of the molecular cloud is enough to absorb X-rays and cause the offset to the right of the observed X-ray peak respect to the likely true peak of the emission.  

HESS J1834-087 has an extended nature as revealed by the MAGIC and HESS observations (Albert et al.~2006; Aharonian et al.~2006).  From the Swift/X-Ray Telescope (XRT) observations of W41, Landi et al.~(2006) found a faint X-ray source within the extent of HESS J1834-087 and thereby suggested a possible pulsar wind nebula (PWN) association. This X-ray source, located at ({\sl l}, {\sl b}) = (23.2340,-0.2657), is also detected by the present XMM-Newton observation,
but there is no evidence for the existence of a PWN. Instead, the prominent diffuse X-ray feature is most likely associated with the TeV $\gamma$-ray emission. 

Yamazaki et al.~(2006) studied the X-ray and $\gamma$-ray emission from evolved SNRs with an age of around 10$^5$ yr and that from a GMC interacting with the SNR. They showed that TeV $\gamma$-ray emission can originate from the SNR, or from the SNR shock running into a GMC, or from the GMC illuminated by high energy protons from the SNR shock. These different origins may be distinguished by the X-ray to $\gamma$-ray spectra. A simple diagnostic is the ratio of the $\gamma$-ray to the X-ray flux, $R_{TeV/X}=F_{\gamma}(1-10{\rm~TeV})/F_X(2-10{\rm~keV})$. 
According to Yamazaki et al.~(2006), for the three cases (in the above mentioned order), the value of $R_{TeV/X}$ is of order $10-10^2$, 10 and $>10^2$, respectively, for their fiducial SNR parameters. For young SNRs found to show gamma-ray emission, $R_{TeV/X}$ is typically below $\sim$2. For HESS J1834-087, the 1-10 TeV $\gamma$-ray flux is $\sim8{\times}10^{-12}{\rm~ergs~s^{-1}~cm^{-2}}$,
based on the MAGIC observation (Albert et al.~2006). Our 2-10 keV X-ray flux is $\sim7{\times}10^{-13}{\rm~ergs~s^{-1}~cm^{-2}}$ giving $R_{TeV/X}\simeq11$, with uncertainty of a factor of two. HESS J1834-087 is possibly arising from a shocked GMC. This is supported by the presence of the GMC as found from the $^{13}$CO detection. The velocities indicate that the GMC is just behind W41. The absorption column density of a few 10$^{22}$ cm$^{-2}$ estimated from the X-ray spectral fit (Table~\ref{tab:spec}) also implies that the nonthermal X-ray emission associated with HESS J1834-087 arises from within the GMC and is behind W41. 
According to Yamazaki et al.~(2006), the X-ray emission from a shocked GMC or a GMC illuminated by high energy protons
is likely synchrotron emission
from secondary electrons arising from hadronic processes. The hard X-ray spectrum that we find is also consistent with such a scenario.

Pulsar J1833-0827 (b=23.386$^{0}$, l=0.063$^{0}$) lies at the north side and about 10$^{\prime}$ away from edge of W41. It has a kinematic distance of 4-5 kpc by HI absorption (Weisberg et al. 1995) and a DM distance of 5.7 kpc and a characteristic age of 147 kyr (Taylor et al. 1993). 
Our results for distance (4 kpc) and age ($\sim$10$^{5}$ yrs) of W41 are consistent with the pulsar J1833-0827, and support the possible associations W41/PSR J1833-0827 also. However, the pulsar is about 20$^{\prime}$ away from the extended $\gamma$-ray source and is not associated with HESS J1834-087. 

\begin{acknowledgements}
TWW and LDA acknowledge support from the Natural Sciences and Engineering Research Council of Canada.  The research at UMass is supported by the NASA/CXC under the grant GO5-6057X. TWW thanks support from the Natural Science Foundation of China, and Dr. Kothes for helpful discussion. We thank Dr. Stil for providing the VGPS data. 
This publication makes use of molecular line data from the Boston University-FCRAO Galactic Ring Survey (GRS). 
The NRAO is a facility of the National Science Foundation operated under cooperative agreement by Associated Universities, Inc. 
\end{acknowledgements}

\clearpage

\begin{deluxetable}{ll}
\tablecaption{Fit to the X-ray spectrum$^a$}
\tablewidth{0pt}
\tablehead{
\colhead{Parameter} &
\colhead{Value}}
\startdata
$\chi^2/d.o.f.$ \dotfill & 61.3/74 \\
$N_{\rm H,1}$ ($10^{20}{\rm~cm^{-2}}$) \dotfill & 3.0 ($<$9.0)$^b$ \\
Temperature (keV; APEC) \dotfill & 0.09$^{+0.03}_{-0.02}$   \\
Photon index (PL1) \dotfill & 0.80$^{+0.13}_{-0.11}$ \\
$N_{\rm H,2}$ ($10^{22}{\rm~cm^{-2}}$) \dotfill & 6.2$^{+3.1}_{-2.5}$\\
Photon index (PL2) \dotfill &  2.0$^{+0.7}_{-0.8}$\\
Flux ($10^{-13}{\rm~ergs~s^{-1}~cm^{-2}}$)$^c$ \dotfill & 7.0 \\
\enddata
\tablerefs{
 $^a$An combined model of absorbed ($N_{\rm H,1}$) APEC+power-law (PL1) is used for characterizing the sky background
and a second absorbed ($N_{\rm H,1}$) power-law (PL2) is used to model to the emission from the feature;
 $^b$Quoted uncertainties are at 90$\%$ confidence level;
 $^c$2-10 keV intrinsic flux for PL2.
}
\label{tab:spec}
\end{deluxetable}

\clearpage

\begin{figure}
\plotone{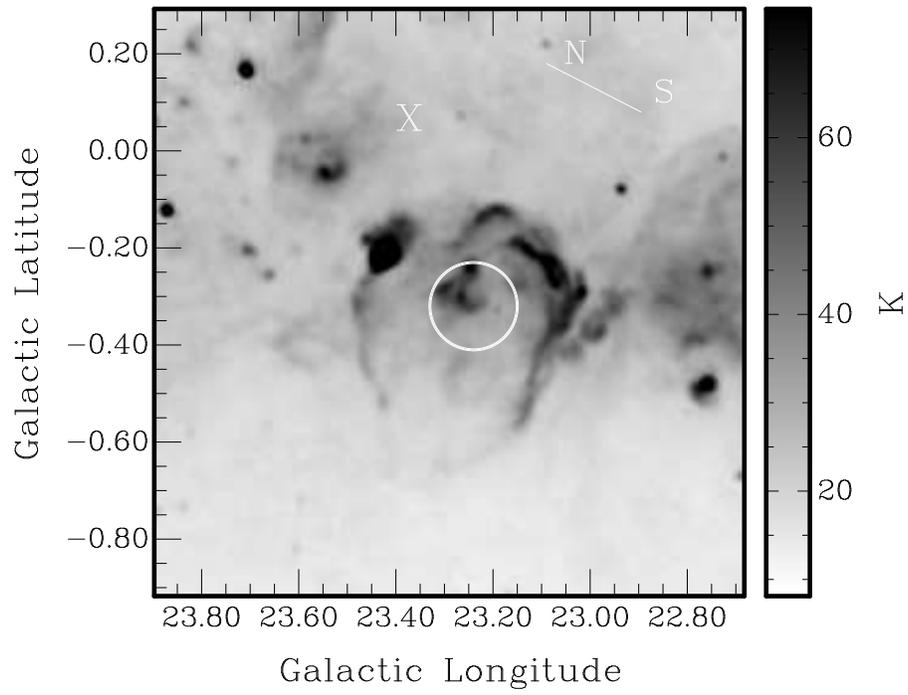}
\caption{The VGPS continuum image of W41 at 1420 MHz. The central circle shows position and extent of the TeV $\gamma$-ray HESS J1934-087. The pulsar PSR J1933-0827 is marked by letter X. The direction of North(N) and South(S) is marked.}
\end{figure}

\clearpage

\begin{figure}
\plotone{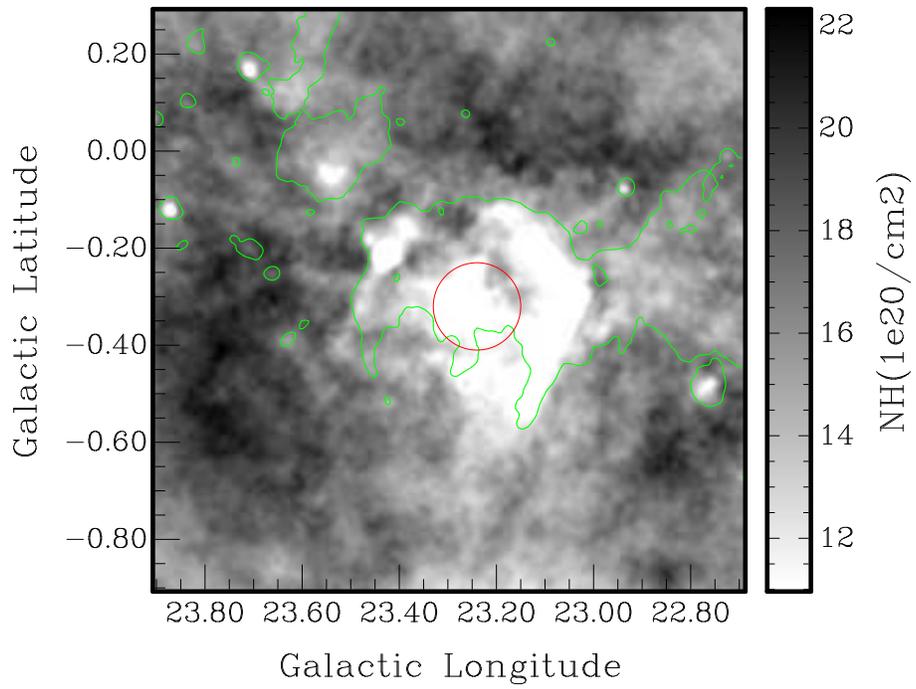}
\caption{The the column density map of the VGPS HI-line emissions associated with W41. This map has superimposed on a W41's contour at 30 K of continuum emission at 1420 MHz chosen to show the SNR (green). The red circle shows the position and extent of HESS J1834-087}
\end{figure}

\clearpage

\begin{figure}
\plotone{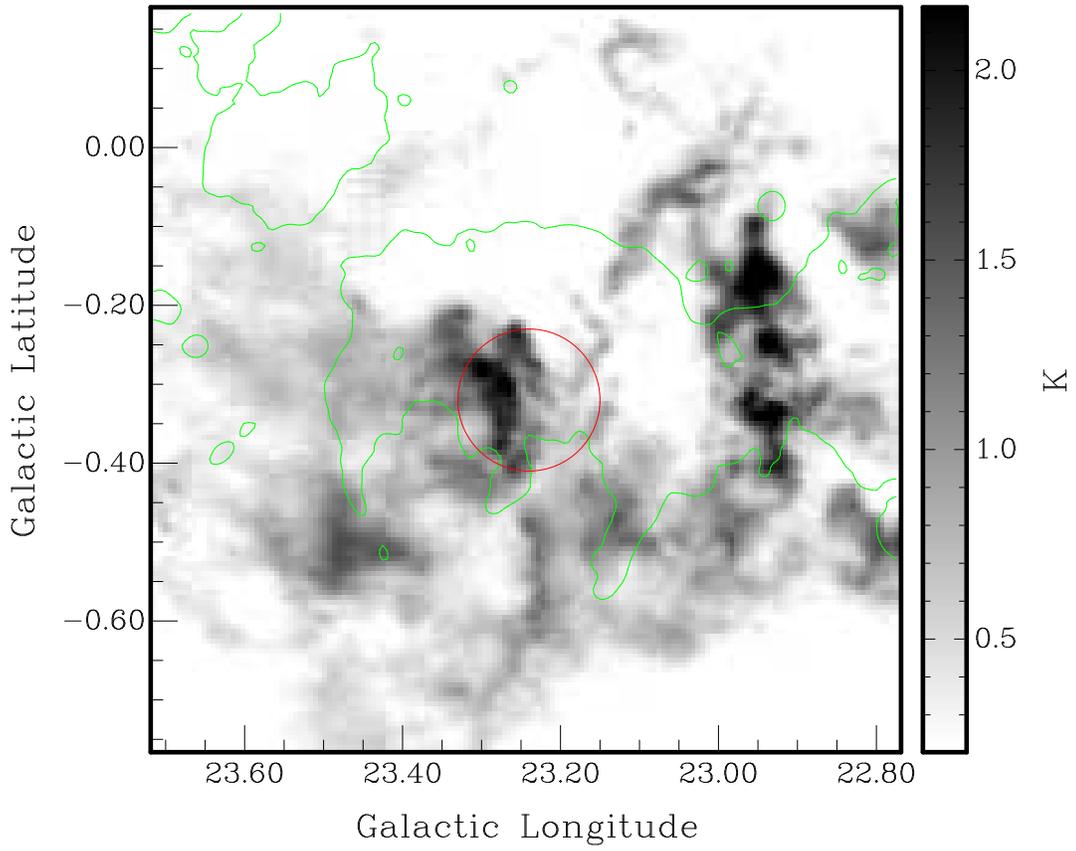}
\caption{The averaged $^{13}$CO emission in the field centered on W41 from 61 to 66 km$/$s. The contour (green) and circle (red) in the map have same meaning as Fig. 2}
\end{figure}

\clearpage

\begin{figure}
\plottwo{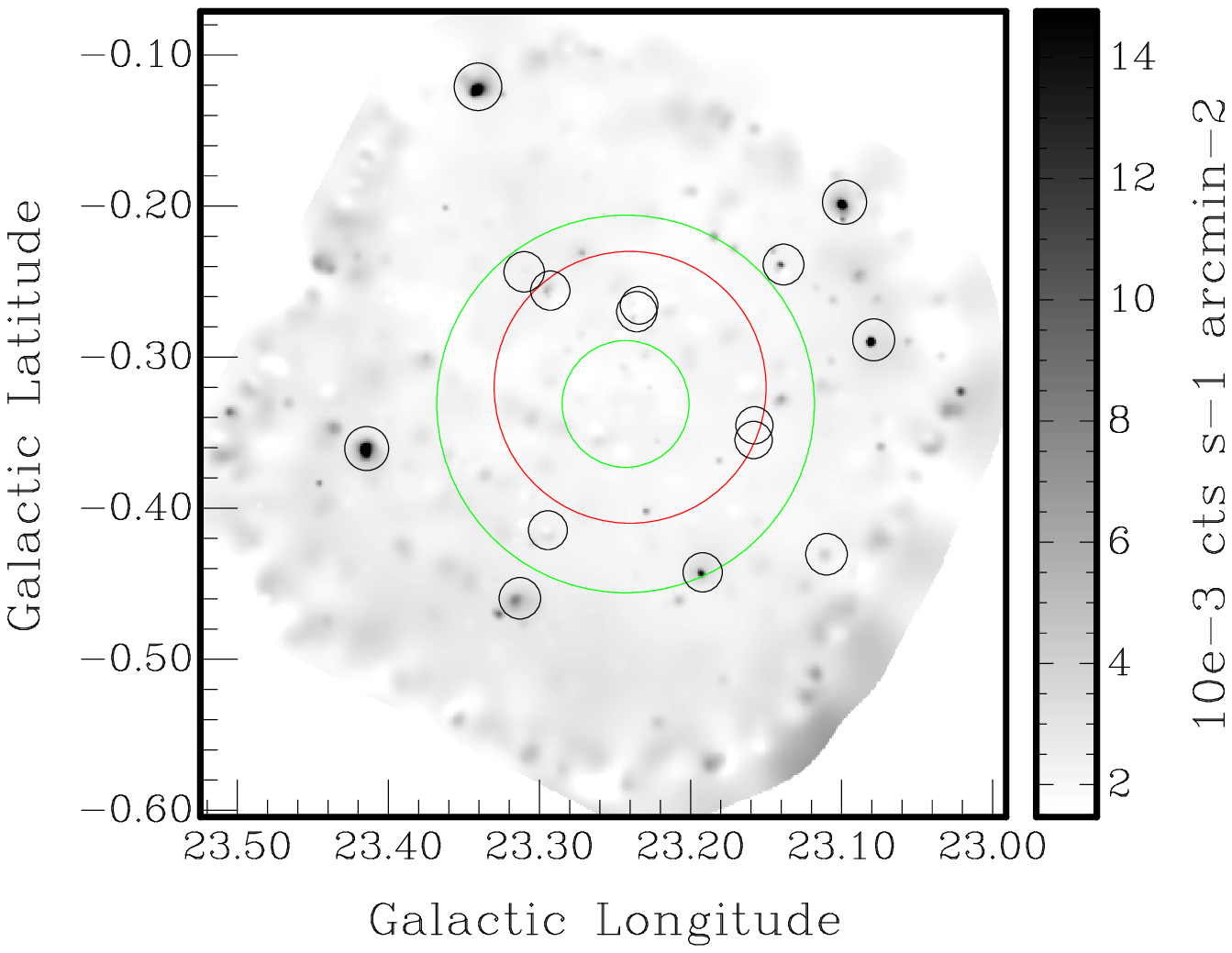}{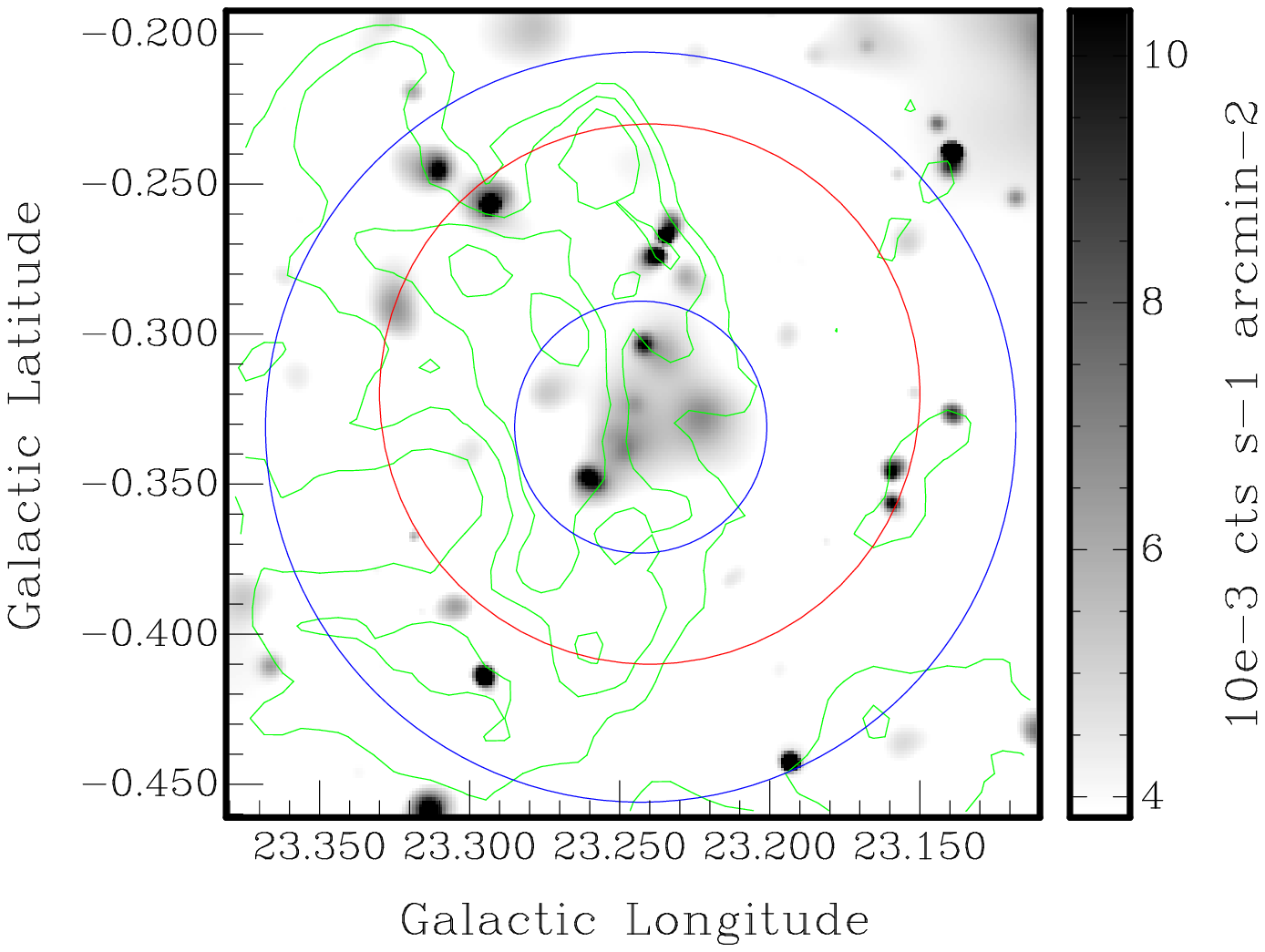}
\caption{{\sl XMM-Newton} EPIC-PN intensity images of SNR W41
in the 0.3-1.5 ({\sl left}) and 1.5-7 keV ({\sl right}) bands.
The intensity is adaptively smoothed using {\sl csmooth} to achieve a signal-to-noise ratio of $\sim$3. 
The middle solid circle (red) represents the location and extent of the $\gamma$-ray source HESS J1834-087. The small circle (blue) illustrates the region of spectral interest, while the large circle (blue) outlines the region (with the enclosed middle circle excluded) where the background spectrum is extracted.
The smaller solid circles (grey in the left plot) outline detected sources and their extent according to twice the 50$\%$ encircled energy radius, which are excluded from spectral extraction. The right plot has superimposed on contours of $^{13}$CO emissions from Fig. 3, at 0.8, 1.2, 1.6 and 2.1 k}
\label{fig:w41_x}
\end{figure}

\clearpage

\begin{figure}
\plotone{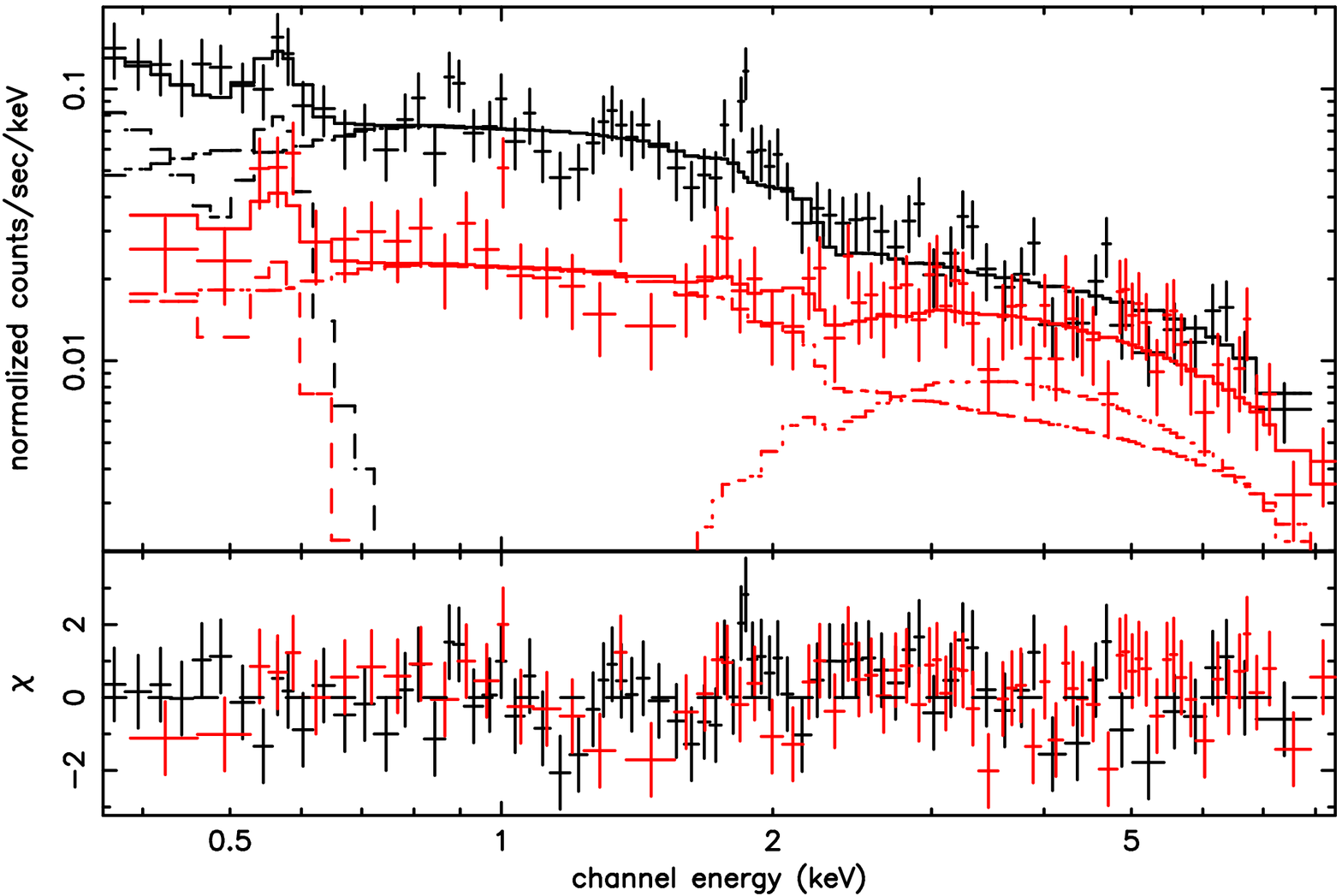}
\caption{EPIC-PN spectra from the SNR W41 region. {\sl Red}: the spectrum of the X-ray feature
associated with HESS J1834-087; {\sl Black}: the spectrum of local sky background. The spectra are
instrumental background-subtracted and are binned to achieve a signal-to-noise ratio better than 3.
Also shown are the best-fit models (solid curves) to the spectra and the different model components (dashed and dotted curves). See text for details }
\label{fig:w41_spec}
\end{figure}


\clearpage

\begin{thebibliography}{}
\bibitem[2006]{Ahaet06}Aharonian, F., Akhperjanian, A.G., Bazer-Bachi, A.R. et al., 2006, ApJ, 636, 777
\bibitem[2005]{Ahaet05}Aharonian, F., Akhperjanian, A.G., Bazer-Bachi, A.R. et al. 2005a, A$\&$A, 437, L7
\bibitem[2005]{Ahaet04}Aharonian, F., Akhperjanian, A.G., Aye, K.-M. et al. 2005b, Science, 307, 1938
\bibitem[2006]{Albet06}Albert, J., Aliu, E., Anderhub, H. et al. 2006, ApJ, 643, L53 
\bibitem[1998]{Comet98}Combi, J.A., Romero, G.E., Benaglia, P., 1998, A\&A, 333, 91L
\bibitem[1972]{Coxet72}Cox, D., 1972, ApJ, 178, 159
\bibitem[2001]{Damet01}Dame, T.M., Hartmann, D., Thaddeus, P., 2001, ApJ, 547, 792	
\bibitem[1986]{Damet86}Dame, T.M., Elmegreen, B.G., Cohen, R.S. et al., 1986, ApJ, 305, 892
\bibitem[2002]{Enoet02}Enomoto R., Tanimori, T., Naito, T. 2002, Nature, 416, 823
2004, MNRAS, 353, 1311
\bibitem[2006]{Lanet04}Landi, R., Bassani, L., Malizia, A., 2006, ApJ, 651, 190 
\bibitem[1990]{Lanet90}Langer, W.D., Penzias, A.A., 1990, ApJ, 357, 477
 \bibitem[1995]{Gaeet95}Gaensler B.M.$\&$ Johnston, S., 1995, MNRAS, 275, 73
\bibitem[1992]{Kaset92}Kassim, N.E., 1992, AJ, 103, 943
\bibitem[2006]{Jaket06}Jackson, J.M., Rathborne, J.M., Shah, R.H. et al. 2006, ApJ Suppl., 163, 145 
\bibitem[2001]{Malet01}Malkov E., Drury L.O'C., 2001, Rep. Prog. Phys., 64, 429
\bibitem[1993]{Tayet93}Taylor, J.H., Manchester, R.N. $\&$ Lyne, A.G., 1993, ApJS, 88, 529.
\bibitem[2003]{Toret03}Torres, D.F., Romero, G.E., Dame, T.M. et al. 2003, Phy. Reports, 382, 303
\bibitem[2006]{Stiet06}Stil, J.M., Taylor, A.R., Dickey, J.M. et al. 2006, AJ, 132, 1158
\bibitem[2004]{Wanet04}Wang Q.D., 2004, ApJ, 612, 159
\bibitem[1996]{Weeet96}Weekes, T.C., Cawley, M.F., Fegan, D.J. et al., 1989, ApJ, 342, 379
\bibitem[1995]{Weset95}Weisberg J.M., Siegel, M.H., Frail, D.A. et al., 
1995, 447, 204
\bibitem[2006]{Yamet06}Yamazaki, R., Kohri, K., Bamba, A. et al. 2006, MNRAS, 371, 1975
\end{thebibliography}
\end{document}